\documentstyle[preprint,tighten,aps]{revtex}
\begin{document} 
\draft
\preprint{IASSNS-HEP-96/27,PUPT-1605}
\date{March 1996}
\title{Entropy of Non-Extreme  Charged Rotating  Black Holes in 
String Theory}
\author{Mirjam Cveti\v c$^1$
\thanks{On sabbatic leave from the University of Pennsylvania.
E-mail address: cvetic@sns.ias.edu}
and Donam Youm$^2$
\thanks{On leave from the University of Pennsylvania. 
E-mail addresses: youm@pupgg.princeton.edu; youm@sns.ias.edu}}
\address{$^1$ School of Natural Science, Institute for Advanced
Study\\ Olden Lane, Princeton, NJ 08540 \\ and \\
$^2$ Physics Department, Joseph Henry Laboratories\\ 
Princeton University, Princeton, NJ 08544}
\maketitle
\begin{abstract}
{We give the explicit expression for four-dimensional rotating charged 
black hole solutions of $N=4$ (or $N=8$) superstring 
vacua, parameterized by the ADM mass, four charges (two 
electric and two magnetic charges, each arising from a different 
$U(1)$ gauge factors), and the angular momentum (as well as the 
asymptotic values of four toroidal moduli of two-torus and 
the dilaton-axion field).  The explicit form of the thermodynamic 
entropy is parameterized in a suggestive way as a sum of the 
product of the `left-moving' and the `right-moving' terms, 
which may have an interpretation in terms of the microscopic 
degrees of freedom of the corresponding $D$-brane configuration.  
We also give an analogous parameterization of the thermodynamic 
entropy for the recently obtained  five-dimensional rotating 
charged black holes parameterized by the ADM mass, three $U(1)$ 
charges and two rotational parameters (as well as the asymptotic 
values of one toroidal modulus and the dilaton).}
\end{abstract}

\section{Introduction}

String theory has reached an exciting stage, where it has now  
become possible to address the long standing problems of 
(quantum) gravity, {\it i.e.}, the microscopic origin of the 
black hole entropy and possibly the issues of the black hole 
information loss for certain classical black hole solutions of 
effective string theory, whose charges can be identified with 
the Ramond-Ramond (R-R) charges of Type II string theory.  
In this case the black hole configurations can be identified 
with particular $D$-brane configurations, whose microscopic 
degrees can be calculated by applying the `$D$-brane technology'
\footnote{For a review on $D$-brane physics see Ref.\cite{CJP}.}. 
In particular, the microscopic entropy of certain five-dimensional  
BPS-saturated static \cite{SV,CM,DVV} and rotating \cite{BMPV} 
as well as certain four-dimensional BPS-saturated static 
\cite{MSTR,JKM} black holes of $N=4$ (or $N=8$) superstring vacua 
has been calculated in that manner
\footnote{An earlier complementary approach to calculate the 
microscopic entropy of four-dimensional BPS-saturated black holes  
was initiated in Ref. \cite{LW} and further elaborated on in Refs.
\cite{CT,T}.}.  
In addition, the microscopic entropy for certain infinitesimal  
deviations from the BPS-saturated limit for static \cite{CM,HS} 
and rotating \cite{BLMPSV} five-dimensional black holes has also  
been provided.  A reliable microscopic calculation in 
terms of the $D$-brane configurations is possible only in the 
coupling regime where the classical black hole description is not 
valid and vice versa \cite{SV}.  However, the topological arguments 
(barring (unlikely) phase transitions) for the microscopic 
calculation and the protection from quantum corrections 
(due to the large enough supersymmetry) for classical results 
of the BPS-saturated states allow one to extrapolate \cite{SV}  
the two results in the regime of each other's validity and  
the agreement between them has been obtained.

Incidentally, the `$D$-brane technology' allows one to 
calculate the microscopic entropy of certain types of black 
holes whose explicit (general) classical configurations have 
been constructed only a posteriori.  In particular,
five-dimensional (generating) solutions parameterized 
by the ADM mass, {\it three} $U(1)$ charges \cite{HMS} 
and {\it two} rotational parameters \cite{CYIII} have been obtained
only most recently.  Their BPS-saturated limit \cite{T}, as 
well as specific BPS-saturated \cite{SV,BMPV} and non-extreme
\cite{HS,BLMPSV} solutions with special charge assignments, have 
also been provided only recently.  We should however point 
out that in the case of special charge assignments (along with the  
subsequent rescaling of the asymptotic value(s) of the scalar 
field(s)) the global space-time properties of the solution remain 
the same as in the case of taking all the three charges different.

The situation for classical solutions of four-dimensional black holes 
of $N=4$ string vacua is somewhat better.  General non-extreme 
static dyonic charged black hole solutions \cite{CYII,JMP} were  
given and their BPS-saturated limit was understood \cite{CT}, 
prior to the realization that these solutions have 
an interpretation as  $D$-brane configurations.  
In particular, a dyonic BPS-saturated solution with four different 
charges \cite{CY} turns out to have a suitable parameterization 
in terms of the corresponding $D$-brane configuration, whose 
microscopic entropy has just been calculated \cite{MSTR,JKM}.

As for general non-extreme solutions, the classical solutions are 
believed to suffer from quantum corrections and the microscopic 
calculation of the entropy need not match the classical result.  
However, in Ref. \cite{HMS} the classical entropy of the non-extreme  
five-dimensional static black hole solutions with three charges
has been written in a suggestive manner, {\it i.e.}, as a product  
of three terms,  where each of them may have an interpretation 
in terms of the square root of  numbers of $D$-brane and 
anti-$D$-brane configuration(s).  
This expression coincides with the microscopic calculations of the 
entropy in special limits, but may hold \cite{HMS} in general.
For four-dimensional four-charge solutions  
the classical entropy is also known \cite{CYI}, and when written 
in an analogous form it may also have an interpretation in terms 
of the number of the $D$-brane configurations, as announced 
\cite{HLM} at the end of Ref. \cite{HMS}. 

The purpose of this paper is to obtain the explicit form of the 
thermodynamic black hole entropy for general non-extreme 
{\it rotating} charged black hole solutions (of $N=4$ or $N=8$ 
string vacua) in five as well as four dimensions.  For that 
purpose we present the explicit form of the four-dimensional 
non-extreme rotating charged black hole solution, parameterized by 
the ADM mass, four different $U(1)$ charges and one rotational 
parameter.  We present the explicit form of the thermodynamic 
entropy, which is written in a suggestive way as a sum of the 
`left-moving' and the `right-moving' terms, that may have an 
interpretation in terms of the  degrees of freedom of the 
$D$-brane configuration.  
We shall also present an analogous expression for the classical 
entropy of recently constructed non-extreme five-dimensional 
solution \cite{CYIII} parameterized by the ADM mass, 
three $U(1)$ charges and two rotational parameters.

The paper is organized in the following way.
In Section II the explicit form of the four-dimensional rotating 
charged solution and the thermodynamic entropy with a discussion 
of different limits is given.  In Section III we write down the  
physical parameters and present an analogous form for the entropy 
of the five-dimensional rotating solution.
In Section IV we comment on a potential interpretation of the 
microscopic entropy.

\section{Four-Dimensional Rotating Solution}

We shall present an explicit form of the (generating) solution for 
the four-charge rotating black hole solution of four-dimensional 
$N=4$ (or $N=8$) superstring vacua.  We choose to parameterize the 
generating solution in terms of the massless fields of the heterotic 
string compactified on a six-torus (or Neveu-Schwarz-Neveu-Schwarz 
(NS-NS) sector of the Type IIA string compactified on $T^6$).  
This solution has an equivalent parameterization (due to the 
string-string duality) in terms of the  NS-NS fields of 
Type IIA compactified on $K3\times T^2$ or $T$-dualized Type IIB 
string.  Due to the $T$-duality (or $U$-duality) of the Type IIA 
string, the solutions parameterized in terms of the NS-NS charges 
have a map onto Ramond-Ramond (R-R) charges and thus an 
interpretation in terms of $D$-brane configurations.

\subsection{Effective Action of Heterotic String on Tori}

For the sake of completeness we briefly summarize the 
results of the effective action of toroidally compactified 
heterotic string in $D$-dimensions, following  Refs. 
\cite{MSCH,SENFOUR}.  

The compactification of the extra $(10-D)$ spatial coordinates 
on a $(10-D)$-torus can be achieved by choosing the following 
Abelian Kaluza-Klein Ansatz for the ten-dimensional metric
\begin{equation}
\hat{G}_{MN}=\left(\matrix{e^{a\varphi}g_{{\mu}{\nu}}+
G_{{m}{n}}A^{(1)\,m}_{{\mu}}A^{(1)\,n}_{{\nu}} & A^{(1)\,m}_{{\mu}}
G_{{m}{n}}  \cr  A^{(1)\,n}_{{\nu}}G_{{m}{n}} & G_{{m}{n}}}\right),
\label{4dkk}
\end{equation}
where $A^{(1)\,m}_{\mu}$ ($\mu = 0,1,...,D-1$; 
$m=1,...,10-D$) are $D$-dimensional Kaluza-Klein $U(1)$ gauge fields,  
$\varphi \equiv \hat{\Phi} - {1\over 2}{\rm ln}\,{\rm det}\, G_{mn}$ 
is the $D$-dimensional dilaton field, and $a\equiv {2\over{D-2}}$.   
Then, the affective action is specified by the following massless 
bosonic fields: the (Einstein-frame) graviton $g_{\mu\nu}$, the 
dilaton $e^{\varphi}$, $(36-2D)$ $U(1)$ gauge fields ${\cal A}^i_{\mu}
\equiv (A^{(1)\,m}_{\mu},A^{(2)}_{\mu\,m},A^{(3)\,I}_{\mu})$ defined 
as $A^{(2)}_{\mu\,m} \equiv \hat{B}_{\mu m}+\hat{B}_{mn}
A^{(1)\,n}_{\mu}+{1\over 2}\hat{A}^I_mA^{(3)\,I}_{\mu}$, 
$A^{(3)\,I}_{\mu} \equiv \hat{A}^I_{\mu} - \hat{A}^I_m 
A^{(1)\,m}_{\mu}$, and the following symmetric $O(10-D,26-D)$ 
matrix of the scalar fields (moduli):
\begin{equation}
M=\left ( \matrix{G^{-1} & -G^{-1}C & -G^{-1}a^T \cr 
-C^T G^{-1} & G + C^T G^{-1}C +a^T a & C^T G^{-1} a^T 
+ a^T \cr -aG^{-1} & aG^{-1}C + a & I + aG^{-1}a^T} 
\right ), 
\label{modulthree}
\end{equation} 
where $G \equiv [\hat{G}_{mn}]$, $C \equiv [{1\over 2}
\hat{A}^{(I)}_{{m}}\hat{A}^{(I)}_{n}+\hat{B}_{mn}]$ and 
$a \equiv [\hat{A}^I_{{m}}]$ are defined in terms of the 
internal parts of ten-dimensional fields.  Then the effective 
$D$-dimensional effective action takes the form:
\begin{eqnarray}
{\cal L}&=&{1\over{16\pi G_D}}\sqrt{-g}[{\cal R}_g-{1\over (D-2)}
g^{\mu\nu}\partial_{\mu}\varphi\partial_{\nu}\varphi+{1\over 8}
g^{\mu\nu}{\rm Tr}(\partial_{\mu}ML\partial_{\nu}ML)-{1\over{12}}
e^{-2a\varphi}g^{\mu\mu^{\prime}}g^{\nu\nu^{\prime}}
g^{\rho\rho^{\prime}}H_{\mu\nu\rho}H_{\mu^{\prime}\nu^{\prime}
\rho^{\prime}} \cr
&-&{1\over 4}e^{-a\varphi}g^{\mu\mu^{\prime}}g^{\nu\nu^{\prime}}
{\cal F}^{i}_{\mu\nu}(LML)_{ij}
{\cal F}^{j}_{\mu^{\prime}\nu^{\prime}}],
\label{effaction}
\end{eqnarray}
where $g\equiv {\rm det}\,g_{\mu\nu}$, ${\cal R}_g$ is the Ricci 
scalar of $g_{\mu\nu}$, and ${\cal F}^i_{\mu\nu} = \partial_{\mu} 
{\cal A}^i_{\nu}-\partial_{\nu} {\cal A}^i_{\mu}$ are the 
$U(1)^{36-2D}$ gauge field strengths.

The $D$-dimensional effective action (\ref{effaction}) is 
invariant under the $O(10-D,26-D)$ transformations ($T$-duality) 
\cite{MSCH,SENFOUR}:
\begin{equation}
M \to \Omega M \Omega^T ,\ \ \ {\cal A}^i_{\mu} \to \Omega_{ij}
{\cal A}^j_{\mu}, \ \ \ g_{\mu\nu} \to g_{\mu\nu}, \ \ \ 
\varphi \to \varphi, \ \ \ B_{\mu\nu} \to B_{\mu\nu}, 
\label{tdual}
\end{equation}
where $\Omega$ is an $O(10-D,26-D)$ invariant matrix, {\it i.e.}, 
with the following property:
\begin{equation}
\Omega^T L \Omega = L ,\ \ \ L =\left ( \matrix{0 & I_{10-D}& 0\cr
I_{10-D} & 0& 0 \cr 0 & 0 &  I_{26-D}} \right ),
\label{4dL}
\end{equation}
where $I_n$ denotes the $n\times n$ identity matrix.

In particular, for $D=4$ the field strength of the one-form 
field is self-dual and the corresponding equations of motion 
and Bianchi identities are invariant under the $SL(2,R)$ 
transformations ($S$-duality) \cite{SENFOUR}:
\begin{equation}
S \to { {{aS+b}\over{cS+d}}},\ \ M\to M ,\ \ g_{\mu\nu}\to 
g_{\mu\nu},\ \ {\cal F}^i_{\mu\nu} \to (c\Psi + d)
{\cal F}^i_{\mu\nu} + ce^{-2\varphi} (ML)_{ij}
\tilde{\cal F}^j_{\mu\nu},
\label{sdual}
\end{equation}
where  $\tilde{\cal F}^{i\,\mu\nu} = {1\over 2\sqrt{-g}} 
\varepsilon^{\mu\nu\rho\sigma}{\cal F}^i_{\rho\sigma}$ and 
$a,b,c,d \in R$ satisfy $ad-bc=1$.  Here, $S\equiv\Psi + 
ie^{-\varphi}$ is a complex scalar field
\footnote{$\Psi$ is the axion which is equivalent to the two-form 
field $B_{\mu\nu}$ through the duality transformation  
$H^{\mu\nu\rho} = -{e^{2\varphi} \over {\sqrt{-g}}}
\varepsilon^{\mu\nu\rho\sigma}\partial_{\sigma}\Psi$.}.  

At the quantum level, the parameters of both $T$- and $S$-duality 
transformations become integer-valued, corresponding to the exact 
symmetry of the perturbative string theory and the conjectured 
non-perturbative symmetry of string theory, respectively.

\subsection{Explicit Solution}

In order to obtain the explicit form of rotating charged solution, 
we employ the solution generating technique, by performing symmetry
transformations on a known solution.  
In particular, we perform four $SO(1,1)\subset O(8,24)$ boosts 
\cite{CYII} on the four-dimensional Kerr solution, specified by 
the ADM mass $m$ and the rotational parameter $l$ (angular 
momentum per unit mass)
\footnote{Within toroidally compactified heterotic string the 
approach to obtain charged solutions from the neutral
one was spelled out in Ref. \cite{SENBH}.  This method was used to 
obtain, {\it e.g.},  general rotating electrically charged 
solutions in four dimensions \cite{SENBH}, higher dimensional 
general electrically charged static solutions \cite{PEET}  
and rotating solutions (with one rotational parameter) 
\cite{HSEN} as well as the general four-dimensional  
static dyonic solutions \cite{CY,JMP}.  Related techniques were 
recently used to obtain a class of five dimensional charged 
(rotating) solutions \cite{HS,BLMPSV,CY}.}.  Here $O(8,24)$ is 
a symmetry of the effective three-dimensional action for
stationary solutions of toroidally compactified heterotic string
\cite{SENTHREE}.  The four boosts $\delta_{e1},\ \delta_{e2}$, 
$\delta_{p1}$ and $\delta_{p2}$ induce two electric 
$Q_2^{(1),(2)}$ and two magnetic charges $P_1^{(1),(2)}$ 
of $U(1)$ gauge fields $A_{\mu 2}^{(1),(2)}$ and 
$A_{\mu 1}^{(1),(2)}$, respectively.  The solution obtained in
that manner is specified by the ADM mass, {\it four} $U(1)$ charges, 
and one rotational parameter
\footnote{A subset of $T$- and $S$-duality transformations, 
{\it i.e.}, $[SO(6)\times SO(22)]/[SO(4)\times SO(20)]\subset 
O(6,22)$ and $U(1)\subset SL(2,R)$ transformations respectively, 
which do not affect the four-dimensional space-time, provides  
51 additional charge parameters, which allow for a general 
solution specified by 28 electric and 28 magnetic charges subject 
to one constraint.  Thus, the generating solution for the most 
general charged rotating solution should be specified by 
{\it five} charge parameters.   In part, due to the technical 
difficulties we postpone the analysis in this case.}.  
In addition, one can subsequently rescale the asymptotic values 
of the dilaton-axion field and the four toroidal moduli of 
two-torus, {\it i.e.}, the scalar fields that vary with spatial 
direction.

Thus, the starting point is the following four-dimensional 
Kerr solution:  
\begin{eqnarray}
ds^2&=& -{{r^2+l^2{\rm cos}^2\theta-2mr}\over{r^2+l^2{\rm cos}^2
\theta}}dt^2 + {{r^2+l^2{\rm cos}^2\theta}\over {r^2+l^2-2mr}}dr^2 
+(r^2+l^2{\rm cos}^2\theta)d\theta^2 \cr
&+&{{{\rm sin}^2\theta}\over{r^2+l^2{\rm cos}^2\theta}}
[(r^2+l^2)(r^2+l^2{\rm cos}^2\theta)+2ml^2r{\rm sin}^2 \theta]
d\phi^2-{{4mlr{\rm sin}^2 \theta}\over{r^2+l^2{\rm cos}^2\theta}}
dtd\phi, 
\label{4dkerr}
\end{eqnarray}
where $m$ is its ADM mass and $l$ is the rotational parameter.
The explicit sequence of the four boost transformations as well 
as technical details of relating the fields of the effective 
three-dimensional action and the four-dimensional fields are 
detailed in Ref. \cite{CYII} (See also Ref. \cite{CYIII}.).

The final expression of the non-extreme dyonic rotating 
black hole solution in terms of the (non-trivial)
four-dimensional bosonic fields is of the following form
\footnote{The four-dimensional Newton's constant is taken to be 
$G_N^{D=4}={1\over 8}$ and we follow the convention of, {\it e.g.}, 
Ref. \cite{MP}, for the definitions of the ADM mass, charges, 
dipole moments and angular momenta.}:  
\begin{eqnarray}
g_{11}&=&{{(r+2m{\rm sinh}^2 \delta_{p2})(r+2m{\rm sinh}^2 \delta_{e2})
+l^2{\rm cos}^2 \theta}\over {(r+2m{\rm sinh}^2 \delta_{p1})(r+2m
{\rm sinh}^2 \delta_{e2})+l^2{\rm cos}^2\theta}},  \cr
g_{12}&=&{2ml{\rm cos}\theta({\rm sinh}\delta_{p1}{\rm cosh}\delta_{p2}
{\rm sinh}\delta_{qe1}{\rm cosh}\delta_{e2}-{\rm cosh}\delta_{p1}
{\rm sinh}\delta_{p2}{\rm cosh}\delta_{e1}{\rm sinh}\delta_{e2})\over 
{(r+2m{\rm sinh}^2 \delta_{p1})(r+2m{\rm sinh}^2 \delta_{e2})+
l^2{\rm cos}^2\theta}},  \cr
g_{22}&=&{{(r+2m{\rm sinh}^2 \delta_{p1})(r+2m{\rm sinh}^2 \delta_{e1})
+l^2{\rm cos}^2 \theta}\over {(r+2m{\rm sinh}^2 \delta_{p1})(r+2m
{\rm sinh}^2 \delta_{e2})+l^2{\rm cos}^2\theta}},  \cr
B_{12}&=&-{{2ml{\rm cos}\theta({\rm sinh}\delta_{p1}{\rm cosh}
\delta_{p2}{\rm cosh}\delta_{e1}{\rm sinh}\delta_{e2}-{\rm cosh}
\delta_{p1}{\rm sinh}\delta_{p2}{\rm sinh}\delta_{e1}{\rm cosh}
\delta_{e2})}\over{(r+2m{\rm sinh}^2 \delta_{p1})(r+2m{\rm sinh}^2 
\delta_{e2})+l^2{\rm cos}^2\theta}},  \cr
e^{\varphi}&=&{{(r+2m{\rm sinh}^2 \delta_{p1})(r+2m{\rm sinh}^2 
\delta_{p2})+l^2{\rm cos}^2 \theta}\over \Delta^{1\over 2}},\cr
ds^2_{E}&=&\Delta^{1\over 2}[-{{r^2-2mr+l^2{\rm cos}^2\theta}\over 
\Delta}dt^2+{{dr^2}\over{r^2-2mr+l^2}} + d\theta^2 
+{{{\rm sin}^2\theta}\over \Delta}\{(r+2m{\rm sinh}^2 
\delta_{p1})\cr 
&\times&(r+2m{\rm sinh}^2 \delta_{p2})(r+2m{\rm sinh}^2 \delta_{e1})
(r+2m{\rm sinh}^2 \delta_{e2})+l^2(1+{\rm cos}^2\theta)r^2+W \cr 
&+&2ml^2r{\rm sin}^2\theta\}d\phi^2
-{{4ml}\over \Delta}\{({\rm cosh} \delta_{p1}{\rm cosh}
\delta_{p2}{\rm cosh} \delta_{e1}{\rm cosh} \delta_{e2} \cr 
&-&{\rm sinh} \delta_{p1}{\rm sinh} \delta_{p2}
{\rm sinh} \delta_{e1}{\rm sinh} \delta_{e2})r
+2m{\rm sinh}\delta_{p1}{\rm sinh}\delta_{p2}{\rm sinh}\delta_{e1}
{\rm sinh}\delta_{e2}\}{\rm sin}^2 \theta dtd\phi],
\label{4dsol}
\end{eqnarray}
where
\begin{eqnarray}
\Delta &\equiv& (r+2m{\rm sinh}^2 \delta_{p1})
(r+2m{\rm sinh}^2 \delta_{p2})(r+2m{\rm sinh}^2 \delta_{e1})
(r+2m{\rm sinh}^2 \delta_{e2})\cr
&+&(2l^2r^2+W){\rm cos}^2\theta, \cr
W &\equiv& 2ml^2({\rm sinh}^2\delta_{p1}+{\rm sinh}^2\delta_{p2}+
{\rm sinh}^2\delta_{e1}+{\rm sinh}^2\delta_{e2})r \cr
&+&4m^2l^2(2{\rm cosh}\delta_{p1}{\rm cosh}\delta_{p2}{\rm cosh}
\delta_{e1}{\rm cosh}\delta_{e2}{\rm sinh}\delta_{p1}{\rm sinh}
\delta_{p2}{\rm sinh}\delta_{e1}{\rm sinh}\delta_{e2}\cr 
&-&2{\rm sinh}^2 \delta_{p1}{\rm sinh}^2 \delta_{p2}{\rm sinh}^2 
\delta_{e1}{\rm sinh}^2 \delta_{e2}-{\rm sinh}^2 \delta_{p2}
{\rm sinh}^2 \delta_{e1}{\rm sinh}^2 \delta_{e2} \cr
&-&{\rm sinh}^2 \delta_{p1}{\rm sinh}^2 \delta_{e1}
{\rm sinh}^2 \delta_{e2}-{\rm sinh}^2 \delta_{p1}{\rm sinh}^2 
\delta_{p2}{\rm sinh}^2 \delta_{e2}-{\rm sinh}^2 \delta_{p1}
{\rm sinh}^2 \delta_{p2}{\rm sinh}^2 \delta_{e1})\cr
&+&l^4{\rm cos}^2 \theta.
\label{4ddef}
\end{eqnarray}
The axion field $a$ also varies with  spatial coordinates, but 
its expression turns out to be cumbersome. 

The ADM mass, $U(1)$ charges $Q^{(1),(2)}_2, 
P^{(1),(2)}_1$, and the angular momentum $J$, can be expressed 
in terms of $m$, $l$ and four boosts in the following way:
\begin{eqnarray}
M&=&4m({\rm cosh}^2 \delta_{e1}+{\rm cosh}^2 \delta_{e2}+
{\rm cosh}^2 \delta_{p1}+{\rm cosh}^2 \delta_{p2})-8m, \cr
Q^{(1)}_2 &=&4m{\rm cosh}\delta_{e1}{\rm sinh}\delta_{e1},\ \ \ \ \
Q^{(2)}_2 = 4m{\rm cosh}\delta_{e2}{\rm sinh}\delta_{e2}, \cr
P^{(1)}_1 &=&4m{\rm cosh}\delta_{p1}{\rm sinh}\delta_{p1},\ \ \ \ \
P^{(2)}_1 = 4m{\rm cosh}\delta_{p2}{\rm sinh}\delta_{p2}, \cr
J&=&8lm({\rm cosh}\delta_{e1}{\rm cosh}\delta_{e2}{\rm cosh}\delta_{p1}
{\rm cosh}\delta_{p2}-{\rm sinh}\delta_{e1}{\rm sinh}\delta_{e2}
{\rm sinh}\delta_{p1}{\rm sinh}\delta_{p2}).  
\label{4dphys}
\end{eqnarray}
The electric dipole moments $D^{(1,2)}_1$ and the magnetic dipole 
moments $\mu^{(1,2)}_2$ of the above solution can also be obtained 
by considering the asymptotic behavior of the solutions 
near spatial infinity: 
\begin{eqnarray}
D^{(1)}_1&=&-4lm({\rm sinh}\delta_{e1}{\rm sinh}\delta_{e2}
{\rm cosh}\delta_{p1}{\rm sinh}\delta_{p2}-
{\rm cosh}\delta_{e1}{\rm cosh}\delta_{e2}{\rm sinh}\delta_{p1}
{\rm cosh}\delta_{p2}), \cr
D^{(2)}_1&=&-4lm({\rm sinh}\delta_{e1}{\rm sinh}\delta_{e2}
{\rm sinh}\delta_{p1}{\rm cosh}\delta_{p2}-
{\rm cosh}\delta_{e1}{\rm cosh}\delta_{e2}{\rm cosh}\delta_{p1}
{\rm sinh}\delta_{p2}), \cr
\mu^{(1)}_2&=&4lm({\rm sinh}\delta_{p1}{\rm sinh}\delta_{p2}
{\rm cosh}\delta_{e1}{\rm sinh}\delta_{e2}-
{\rm cosh}\delta_{p1}{\rm cosh}\delta_{p2}{\rm sinh}\delta_{e1}
{\rm cosh}\delta_{e2}), \cr
\mu^{(2)}_2&=&4lm({\rm sinh}\delta_{p1}{\rm sinh}\delta_{p2}
{\rm sinh}\delta_{e1}{\rm cosh}\delta_{e2}-
{\rm cosh}\delta_{p1}{\rm cosh}\delta_{p2}{\rm cosh}\delta_{e1}
{\rm sinh}\delta_{e2}).   
\label{4mom}
\end{eqnarray}
The solution corresponds effectively to the six-dimensional 
target-space background with the four toroidal moduli of two-torus  
($T^2$) and the dilaton-axion field varying with the spatial 
directions.  

In the above expressions the asymptotic values of four moduli of
$T^2$  are taken to have canonical values 
$g_{11\,\infty}=g_{22\,\infty}=1$, $B_{12\,\infty}=g_{12\,\infty}=0$, 
but can be rescaled (along with the physical charges) by an 
arbitrary $O(2,2)$ constant matrix ({\it cf.}, (\ref{tdual})).  
Also, the canonical choice of the asymptotic value of the 
dilaton-axion field, $\varphi_\infty=a _\infty=0$, can be 
rescaled (along with the physical charges) by an arbitrary 
$SL(2,R)$ constant matrix ({\it cf.}, (\ref{sdual})).
Note, none of  these rescaling transformations changes the 
four-dimensional space-time, only the physical interpretation 
of the charge parameters in (\ref{4dphys}) changes.  We will 
primarily stick to the representation with the canonical choices 
of the asymptotic values of the scalar fields
\footnote{On the other hand, a judicial choice of different 
asymptotic values of scalar fields is useful (see, {\it e.g.}, 
Refs. \cite{HS,BLMPSV}) for calculating reliably the microscopic 
entropy for the infinitesimal deviations from the BPS-saturated 
limits.}.

The solution (\ref{4dsol}) has the inner $r_-$ and the outer $r_+$ 
horizons at:
\begin{equation}
r_{\pm}=m\pm \sqrt{m^2-l^2},
\label{horizon}
\end{equation}
provided $m\ge |l|$.  In this case the solution has the 
global space-time of the Kerr-Newman  black hole, {\it i.e.}, the
rotating charged black hole of the 
Maxwell-Einstein gravity, with the ring singularity at 
$r={\rm min}\{Q^{(1)}_2,Q^{(2)}_2,P^{(1)}_1,P^{(2)}_1\}$ and 
$\theta = {\pi\over 2}$.  When $\delta_{e1}=\delta_{e2}=\delta_{p1}
=\delta_{p2}$, {\it i.e.}, $Q_2^{(1)}=Q_2^{(2)}=P_1^{(1)}=P_1^{(2)}$, 
all the toroidal moduli and the axion-dilaton field are constant,  
and thus the solution {\it is} the Kerr-Newman solution
\footnote{The case with $\delta_{p1}=\delta_{p2}=0$ is a 
generating solution of a general electrically charged rotating 
solution of Ref. \cite{SENBH}.  The case with $\delta_{e1}=
\delta_{p1}$, $\delta_{e2}=\delta_{p2}$, {\it i.e.}, 
$Q_2^{(1)}=P_1^{(1)}$, $Q_2^{(2)}=P_1^{(2)}$, was recently 
constructed in Ref. \cite{JMPII}.}. 

The extreme solution, {\it i.e.}, the case when  the inner and 
the outer horizons coincide, is reached when  $m\to |l|^+$.  
In this case, the global space-time is that of extreme Kerr-Newman 
solution. 

The BPS-saturated limit, {\it i.e.}, when the configuration 
saturates the Bogomol'nyi bound for the ADM mass, is reached 
when $m\to 0$, while the charges $Q_2^{(1),(2)}$ and 
$P_1^{(1),(2)}$, and the angular momentum $J$ are 
kept finite.  For  the charges to 
be constant, the boosts $\delta_{e1,e2,p1,p2}\to \infty$,  
while keeping $m{\rm e}^{2\delta_{e1,e2,p1,p2}}$ constant.  
In order for $J$ to remain {\it non-zero}, the rotational  
parameter $l$ has to remain {\it non-zero}.  Thus, the 
BPS-saturated charged solution with the non-zero angular 
momentum $J$  has a naked singularity
\footnote{The conformal two-dimensional $\sigma$-model, whose 
target space corresponds to dyonic rotating  BPS-saturated 
solutions, also yields solutions with naked singularities 
\cite{TU}.}, 
since the constraint $m\ge |l|$ is not satisfied.
Thus the existence of the naked singularities in the case of 
rotating BPS-saturated solution persists even in the case of 
four-nonzero charges
\footnote{This fact was also known for the case of four-dimensional 
electrically charged solutions \cite{HSEN}, whose static 
BPS-saturated subset has the null horizon.  On the other hand, 
in five dimensions,  the charged solution with three nonzero 
charges has a regular  BPS-saturated limit with the non-zero angular 
momentum \cite{BMPV,T,CY}.  In six-dimensions, the electrically 
charged BPS-saturated rotating solution is also regular \cite{HSEN}.}.
  
Thus, the only regular BPS-saturated solution is in this case the
static solution (with zero angular momentum $J=0$, {\it i.e.}, 
$l\to 0$).  Its global space-time is that of extreme  
Reissner-Nordst\" om black holes.

\subsection{Entropy of the Four-Dimensional Rotating Solution}

The thermodynamic (Bekenstein-Hawking) entropy is of the form
$S=\textstyle{1\over 4G_N} A$, where $A$ is the surface area
$A=\int d\theta d\phi \left.\sqrt{g_{\theta\theta}g_{\phi\phi}}
\right|_{r=r_+}$, determined at the outer-horizon 
$r_+=m+\sqrt{m^2-l^2}$.  In the case of (\ref{4dsol}) 
we were able to cast the thermodynamic entropy  
in the following form:
\begin{eqnarray}
S&=&16\pi [m^2(\prod^4_{i=1}\cosh\delta_i+\prod^4_{i=1}
\sinh\delta_i)+m\sqrt{m^2-l^2}(\prod^4_{i=1}
\cosh\delta_i-\prod^4_{i=1}\sinh\delta_i)]\cr
&=& 16\pi [m^2(\prod^4_{i=1}\cosh\delta_i+\prod^4_{i=1}
\sinh\delta_i)+\{m^4(\prod^4_{i=1}\cosh\delta_i
-\prod^4_{i=1}\sinh\delta_i)^2-J^2\}^{1/2}], 
\label{4dent}
\end{eqnarray}
where $m,\ l $ are again the ADM mass and the angular momentum 
per unit mass of the Kerr solution  (\ref{4dkerr}) and 
$\delta_{1,2,3,4}\equiv \delta_{e1,e2,p1,p2}$ are the four 
boosts specifying the four charges (\ref{4dphys}).  In the 
second line the entropy is cast in terms of the angular 
momentum $J$ of the charged solution. 

Even though one expects the thermodynamic entropy to be cast in a 
form which is a square root of an expression 
(which  depends on  charges and angular momentum),
it can be written in a form (\ref{4dent}) which is a {\it sum} of 
two terms, {\it i.e.},the sum of the `left-moving' and 
the `right-moving' contributions.  Each term is {\it symmetric} 
in terms of the four boost parameters (and thus in terms 
of the four charges).  On the other hand, (\ref{4dent}) 
bears an asymmetry with respect to the angular momentum, 
{\it i.e.},  only the {\it right-moving} term contains  
the angular momentum, which acts as to reduce the right-moving 
contribution to the entropy.

When the angular momentum is zero ($J=0$, {\it i.e.}, $l=0$),
the entropy is of the form \cite{CYI}
\footnote{In Ref. \cite{CYI}, the expression  for the entropy 
 has a typographical error, {\it i.e.}, 
the square root is missing. Also, instead in terms of boosts, the 
expression is given in terms of physical charges and the  
non-extremality parameter $\beta\equiv 2m$.}:
\begin{equation}
S=32\pi m^2\prod^4_{i=1}\cosh\delta_i, 
\end{equation}
which is a generalization of the  static five-dimensional  
result with  three charges \cite{HMS} to the 
static  four-dimensional  result with four-charges. 

In the (regular) BPS-saturated limit ($m \to 0$, $l\to 0$, while
$m{\rm e}^{2\delta_{e1,e2,p1,p2}}$ are kept constant) as well as 
the extreme limit $m\to |l|^+$, the `right-moving' term in 
(\ref{4dent}) is zero, however in terms of physical parameters
(charges and angular momentum), the entropy is different in each
case.  In the BPS-saturated limit:
\begin{equation}
S=32\pi m^2\prod^4_{i=1}\cosh\delta_i=
2\pi[P^{(1)}_1P^{(2)}_1Q^{(1)}_2Q^{(2)}_2]^{1\over 2}, 
\label{4dbpsent}
\end{equation}
while in the extreme limit:
\begin{eqnarray}
S&=&16\pi m^2(\prod^4_{i=1}\cosh\delta_i+\prod^4_{i=1}
\sinh\delta_i)\cr
&=&2\pi[J^2+P^{(1)}_1P^{(2)}_1Q^{(1)}_2
Q^{(2)}_2]^{1\over 2}.
\label{4dexarea}
\end{eqnarray}
Note that in the case when the right-moving contribution is absent, 
both  expressions (\ref{4dbpsent}) and (\ref{4dexarea})
 are independent of the asymptotic values 
of the moduli and the dilaton coupling and have a straightforward 
generalization to the manifestly $S$- and $T$-duality invariant
form\cite{CT}.
Namely, when expressed in terms of the conserved  quantized 
charge electric and magnetic  lattice vectors 
$\vec{\alpha},\vec{\beta} \in \Lambda_{6,22}$ (of toroidally 
compactified heterotic string), the surface area can be written  as 
\begin{equation}
S=2\pi[J^2+\{(\vec{\cal \alpha}^TL\vec{\cal \alpha})
(\vec{\cal \beta}^TL\vec{\cal \beta})-(\vec{\cal \alpha}^TL
\vec{\cal \beta})^2\}]^{1\over 2}. 
\label{4dvecexarea}
\end{equation}

\section{Entropy of the Five-Dimensional Rotating Solution}

In this section we write explicitly the entropy for
five-dimensional rotating charged black holes specified by the ADM 
mass, three charges and two rotational parameters.  
These solutions were obtained \cite{CYIII} by performing three
$SO(1,1)\subset O(8,24)$ boosts, on the five-dimensional (neutral) 
rotating solution parameterized by the mass $m$ and two rotating 
parameters $l_1$ and $l_2$.  The three boosts $\delta_{e1},\ 
\delta_{e2}$, and $\delta_e$ specify respectively the two 
electric charges $Q_1^{(1)},\ Q_1^{(2)}$ of the two $U(1)$ gauge 
fields, {\it i.e.}, the Kaluza-Klein $A_{\mu\,1}^{(1)}$ and the 
two-form $A_{\mu\,1}^{(2)}$ gauge fields associated with the 
first compactified direction, and the charge $Q$, {\it i.e.}, 
the electric charge of the vector field, whose field strength is 
dual to the field strength $H_{\mu\nu\rho}$ of the 
five-dimensional two form field $B_{\mu\nu}$. 

For the sake of completeness we quote the result\cite{CYIII} for 
space-time metric of the solution: 
\begin{eqnarray}
ds^2_E&=& \bar{\Delta}^{1\over 3} \left [-{{(r^2+l^2_1{\rm cos}^2
\theta +l^2_2{\rm sin}^2\theta)(r^2+l^2_1{\rm cos}^2\theta
+l^2_2{\rm sin}^2\theta-2m)}\over \bar{\Delta}} dt^2 \right .
\cr
&+&{r^2 \over {(r^2+l^2_1)(r^2+l^2_2)-2mr^2}} dr^2 +d\theta^2
+{{4m{\rm cos}^2\theta {\rm sin}^2\theta}\over \bar{\Delta}}
[l_1 l_2\{(r^2+l^2_1{\rm cos}^2\theta +l^2_2{\rm sin}^2\theta)
\cr
&-&2m({\rm sinh}^2\delta_{e1}{\rm sinh}^2\delta_{e2}+
{\rm sinh}^2\delta_{e}{\rm sinh}^2\delta_{e1}
+{\rm sinh}^2\delta_{e}{\rm sinh}^2\delta_{e2})\}+2m\{(l^2_1+l^2_2)
\cr
&\times&{\rm cosh}\delta_{e1}{\rm cosh}
\delta_{e2}{\rm cosh}\delta_{e}
{\rm sinh}\delta_{e1}{\rm sinh}\delta_{e2}{\rm sinh}\delta_{e}
-2l_1l_2{\rm sinh}^2 \delta_{e1}{\rm sinh}^2 \delta_{e2}
{\rm sinh}^2 \delta_{e}\}] d\phi d\psi
\cr
&-&{{4m{\rm sin}^2\theta}\over \bar{\Delta}}[(r^2+l^2_1
{\rm cos}^2\theta+l^2_2{\rm sin}^2\theta)(l_1{\rm cosh}
\delta_{e1}{\rm cosh}\delta_{e2}{\rm cosh}\delta_{e}-
l_2{\rm sinh}\delta_{e1}{\rm sinh}\delta_{e2}{\rm sinh}\delta_{e})
\cr
&+&2ml_2{\rm sinh}\delta_{e1}{\rm sinh}\delta_{e2}{\rm sinh}
\delta_{e}]d\phi dt -{{4m{\rm cos}^2\theta}\over\bar{\Delta}}
[(r^2+l^2_1{\rm cos}^2\theta+l^2_2{\rm sin}^2\theta)
\cr
&\times&(l_2{\rm cosh}\delta_{e1}{\rm cosh}\delta_{e2}
{\rm cosh}\delta_{e}-l_1{\rm sinh}\delta_{e1}{\rm sinh}
\delta_{e2}{\rm sinh}\delta_{e})+2ml_1{\rm sinh}\delta_{e1}
{\rm sinh}\delta_{e2}{\rm sinh}\delta_{e}] d\psi dt
\cr
&+&{{{\rm sin}^2\theta}\over \bar{\Delta}}
[(r^2+2m{\rm sinh}^2\delta_e+l^2_1)(r^2+2m{\rm sinh}^2\delta_{e1}
+l^2_1{\rm cos}^2\theta+l^2_2{\rm sin}^2\theta)(r^2+2m{\rm sinh}^2
\delta_{e2} \cr
&+&l^2_1{\rm cos}^2\theta+l^2_2
{\rm sin}^2\theta)+2m{\rm sin}^2 \theta\{(l^2_1{\rm cosh}^2\delta_m
-l^2_2{\rm sinh}^2\delta_m)(r^2+l^2_1{\rm cos}^2\theta +l^2_2{\rm sin}^2
\theta)\cr
&+&4ml_1 l_2{\rm cosh}\delta_{e1}
{\rm cosh}\delta_{e2}{\rm cosh}\delta_{e}{\rm sinh}\delta_{e1}
{\rm sinh}\delta_{e2}{\rm sinh}\delta_{e}-2m{\rm sinh}^2\delta_{e1}
{\rm sinh}^2\delta_{e2} \cr
&\times&(l^2_1{\rm cosh}^2\delta_e+l^2_2{\rm sinh}^2\delta_e)
-2ml^2_2{\rm sinh}^2\delta_e({\rm sinh}^2\delta_{e1}+
{\rm sinh}^2\delta_{e2})\}]d\phi^2
\cr
&+&{{{\rm cos}^2\theta}\over \bar{\Delta}}[(r^2+2m{\rm sinh}^2
\delta_e+l^2_2)(r^2+2m{\rm sinh}^2\delta_{e1}+l^2_1{\rm cos}^2\theta
+l^2_2{\rm sin}^2\theta)(r^2+2m{\rm sinh}^2\delta_{e2}
\cr
&+&l^2_1{\rm cos}^2\theta+l^2_2{\rm sin}^2\theta)
+2m{\rm cos}^2 \theta\{(l^2_2{\rm cosh}^2\delta_e
-l^2_1{\rm sinh}^2\delta_e)(r^2+l^2_1{\rm cos}^2\theta
+l^2_2{\rm sin}^2\theta)
\cr
&+&4ml_1 l_2 {\rm cosh}\delta_{e1}
{\rm cosh}\delta_{e2}{\rm cosh}\delta_{e}{\rm sinh}\delta_{e1}
{\rm sinh}\delta_{e2}{\rm sinh}\delta_{e}-2m
{\rm sinh}^2\delta_{e1}{\rm sinh}^2\delta_{e2}
\cr
&\times&\left .(l^2_1{\rm sinh}^2\delta_e+l^2_2{\rm cosh}^2\delta_e)
-2ml^2_1{\rm sinh}^2\delta_e({\rm sinh}^2\delta_{e1}+{\rm sinh}^2
\delta_{e2})\}]d\psi^2 \right],
\label{5daxisol}
\end{eqnarray}
where
\begin{eqnarray}
\bar{\Delta} &\equiv& (r^2+2m{\rm sinh}^2\delta_{e1}+
l^2_1{\rm cos}^2\theta+l^2_2{\rm sin}^2\theta)(r^2+2m
{\rm sinh}^2\delta_{e2}+l^2_1{\rm cos}^2\theta+l^2_2
{\rm sin}^2\theta)\cr
&\times&(r^2+2m{\rm sinh}^2\delta_{e}+l^2_1{\rm cos}^2\theta
+l^2_2{\rm sin}^2\theta).
\label{5def}
\end{eqnarray}
The  the ADM mass $M$, $U(1)$ charges $Q$'s and the angular momenta  
$J's$ are given in terms of the three boost parameters 
$\delta_{e1,e2,e}$ and the three parameters $m,\ l_1,\ l_2$ of the
neutral rotating solution as
\footnote{The five-dimensional Newton's constant is taken to be 
$G_N^{D=5}={\pi\over 4}$.}:  
\begin{eqnarray}
M&=&2m({\rm cosh}^2\delta_{e1}+{\rm cosh}^2\delta_{e2}+
{\rm cosh}^2\delta_{e})-3m,\cr
Q^{(1)}_1&=&2m{\rm cosh}\delta_{e1}{\rm sinh}\delta_{e1}, \ \ \ \ 
Q^{(2)}_1=2m{\rm cosh}\delta_{e2}{\rm sinh}\delta_{e2}, \ \ \ \ 
Q=2m{\rm cosh}\delta_{e}{\rm sinh}\delta_{e}, 
\cr
J_{\phi}&=&4m(l_1{\rm cosh}\delta_{e1}{\rm cosh}\delta_{e2}
{\rm cosh}\delta_{e}-l_2{\rm sinh}\delta_{e1}{\rm sinh}\delta_{e2}
{\rm sinh}\delta_{e}), \cr
J_{\psi}&=&4m(l_2{\rm cosh}\delta_{e1}{\rm cosh}\delta_{e2}
{\rm cosh}\delta_{e}-l_1{\rm sinh}\delta_{e1}{\rm sinh}\delta_{e2}
{\rm sinh}\delta_{e}).
\label{5ddef}
\end{eqnarray}
Note the solution is effectively the six-dimensional solution 
with the toroidal modulus $g_{11}$  and the dilaton field $\varphi$ 
varying with the spatial coordinates \cite{CY}. 

The solution has the inner $r_-$  and the outer $r_{+} $  
horizons at:
\begin{equation}
r_{\pm}^2=m-{1\over 2}l_1^2-{1\over 2}l_2^2 \pm
{1\over 2}\sqrt{(l^2_1-l^2_2)^2+4m(m-l^2_1-l^2_2)}, 
\label{5dhor}
\end{equation}
provided $2m\ge (|l_1|+|l_2|)^2$.

We write explicitly the classical entropy  
$S={1\over {4G_N}}A$, where $A$ is the surface 
area  $A=\int d\theta d\phi d\psi \left.
\sqrt{g_{\theta\theta}(g_{\phi\phi}g_{\psi\psi}-g_{\phi\psi}^2)}
\right |_{r=r_+}$, determined at the outer-horizon $r_+$.
The entropy can be written in the following form:
\begin{eqnarray}
S&=&4\pi \left[m\{2m- (l_1-l_2)^2\}^{1/2} 
(\prod^3_{i=1}\cosh\delta_i +\prod^3_{i=1}\sinh\delta_i)\right.\cr
& &\ \ \ \ +\left. m\{2m- (l_1+l_2)^2\}^{1/2}(\prod^3_{i=1}\cosh
\delta_i-\prod^3_{i=1}\sinh\delta_i)\right]
\cr
&=&4\pi\left[\{2m^3(\prod^3_{i=1}\cosh\delta_i+\prod^3_{i=1}
\sinh\delta_i)^2-\textstyle{1\over 16}(J_\phi-J_\psi)^2\}^{1/2}
\right.\cr
& &\ \ \ \ +\left.\{2m^3(\prod^3_{i=1}\cosh\delta_i-\prod^3_{i=1} 
\sinh\delta_i)^2-\textstyle{1\over 16}(J_\phi+J_\psi)^2\}^{1/2}
\right],
\label{5dent}
\end{eqnarray}
where $\delta_{1,2,3}\equiv \delta_{e1,e2,e}$  
and  $m$, $l_{1,2}$ are  the ADM mass and the two 
rotational parameters of the five-dimensional rotating
(neutral) solution, respectively.  In the second line, the entropy  
is cast in terms of boosts (specifying the  three charges) and 
the two angular momenta $J_{\psi,\,\phi}$ of the charged solution. 
Again, the classical entropy can  be cast  (\ref{5dent}) 
as  a {\it sum} of two terms, {\it i.e.}, the sum of the 
`left-moving' and the `right-moving' contributions.  

The form of the entropy, as a sum of the two terms, has been 
derived in the case of infinitesimal deviations from the BPS-saturated limit 
in Ref. \cite{BLMPSV}, and its microscopic degrees of freedom were 
identified with the left- and the right-moving contributions of  
the $D-$brane world-volume Hilbert space with $J_{L,R}\equiv 
\textstyle{1\over 2}(J_\phi \mp J_\psi)$ identified as the left- 
(or right-) moving charges of the $U(1)_{L,R}$  
($N=4$) superconformal (world-sheet) algebra. 

Interestingly, even for a general non-extreme solution the 
classical entropy (\ref{5dent}) retains the form as a sum 
of two pieces, one containing  the `left-moving' and another one  the 
`right-moving' contributions, thus suggesting that even for a 
generic non-extreme case the expression may have a microscopic
interpretation in terms of  degrees arising from 
two (left-moving and right-moving) non-interacting 
$D$-brane world-volume sectors.  Note also, that each term 
is {\it symmetric} under the permutation of  the three boost 
parameters and thus under the permutation of the three 
charge assignments.

The (regular) BPS-saturated limit, {\it i.e.}, the limit where 
the ADM mass saturates the Bogomol'nyi bound,  
is reached \cite{CYIII} by taking $m\to 0$, $l_{1,2}\to 0$ and 
$\delta_{e1,e2,e}\to \infty$, while $Q_{1,2}={1\over 2}m 
{\rm e}^{\delta_{e1,e2}}$, $Q={1\over 2}m{\rm e}^{\delta}$ 
and $l_{1,2}/m^{1/2}$ are kept finite.  In this 
case, the right-moving contribution disappears.
Interestingly,   in the extreme limit (the inner and outer
horizons (\ref{5dhor}) coinciding), which corresponds to the choice  
$2m\to (l_1+l_2)^2$,  the right-moving contributions again 
disappears, however, the actual value of the entropy in
terms of the physical parameters is different from the 
BPS-saturated limit.

For zero angular momentum, the entropy formula again rearranges 
itself as a single term \cite{HMS}:
\begin{equation}
S=8\sqrt{2}\pi m^{3/2}\prod^3_{i=1}\cosh\delta_i, 
\end{equation}
being fully symmetric under permutations of  charges.  In this 
case, the microscopic entropy can be calculated 
in certain limits, but it was pointed out \cite{HMS} that its  
validity as a microscopic entropy may be true in general and that 
each (`dressed') boost ${\rm e}^{\delta_i}$ [${\rm e}^{-\delta_i}$] 
may be interpreted as a square root of the number of the 
corresponding $D$-brane [anti-$D$-brane] configurations.
 
A more general expression for the entropy (\ref{5dent}) has a
suggestive form indicating that the relevant charge 
degrees of freedom should be identified with the left- 
and  right-moving ($D$-brane world-volume) sectors, which appear
in combinations $(\prod^3_{i=1}\cosh\delta_i\pm\prod^3_{i=1}
\sinh\delta_i)$, 
respectively.

\section{Comments on the $D$-Brane Interpretation}

We obtained the explicit forms of the classical entropy for 
the four- and five-dimensional rotating charged black hole 
solutions with four charges and one rotational parameter, and three
charges and two rotational parameters, respectively.  These solutions 
can be viewed as `generating' solutions for black holes of $N=4$ 
(or $N=8$) string vacua.  

Even though  we chose to parameterize the classical 
solutions  in  terms fields of the toroidally compactified 
heterotic sting [or equivalently in terms of the NS-NS sector 
fields of the toroidally compactified Type IIA sting], these 
solutions map, due to 
string-string duality [or $U$-duality], onto configurations  
with R-R charges of Type IIA string compactified on $K3\times T^2$ 
[or R-R charges of Type IIA string compactified on $T^6$].  
Thus, they have an interpretation in terms of the 
(intersecting) $D$-brane configuration.

Interestingly, even-though  non-extreme  classical 
solutions may  receive quantum corrections, for  both 
 four-dimensional and the five-dimensional solutions, 
the classical  entropy  (\ref{4dent}) and (\ref{5dent}) 
is  given as  a sum of the `left-moving' and the `right-moving' 
contributions, which is suggestive of a microscopic interpretation 
in terms of  two contributions arising from the (non-interacting) 
left-moving and right-moving  sector of the (intersecting) 
$D$-brane world-volume Hilbert spaces. 

There is an interesting parallel between  the structures of the  
entropy of the four-dimensional (\ref{4dent}) and 
the five-dimensional (\ref{5dent}) solutions.  
The effect of the fourth charge in four dimensions is  an additional
 factor (in the products) associated with the fourth boost 
({\it i.e.}, the fourth charge).  In either case, the 
expression is fully symmetric under permutations of 
boosts (charges). On the other hand, in going from 
five to four dimensions  the left-moving angular momentum disappears, 
while the right-moving angular momentum effectively remains non-zero,  
as if the limit $J_\psi\to J_\phi$ is taken.  From the microscopic 
point of view, this suggests that in the case of rotating 
four-dimensional configurations  the $D$-brane world-volume  
(world-sheet) is specified by the ($N=2$) superconformal algebra 
of the right-moving sector, only, and therefore the states 
are identified under the $U(1)_R$ superconformal currents, only.

Since the structure of the classical entropy (\ref{4dent}) 
[or (\ref{5dent})] suggests a  full symmetry among the four 
[or the three] charges, it may be preferable to identify the  
(intersecting) $D$-brane configuration of four [or three]-different 
types of $D$-branes \cite{CS} whose world-volume excitations would 
account for the different charge degrees of freedom in a symmetric
way \footnote{Note, however, that this is not the approach used in 
recent calculations of the microscopic entropy.}.   
In particular, the four-dimensional static generating solution  
of $N=4$ [or $N=8$] superstring vacua can be interpreted \cite{CS} 
(in terms of the Type IIA string) as an intersecting $D$-brane
 configuration of 
$Q_2^{(1)}$ zero-branes, and $Q_2^{(2)}$, $P_1^{(1)}$ and $P_1^{(2)}$ 
four-branes wrapping around $K3$, $S_1^2\times T^2$ and $S_2^2\times 
T^2$ [or wrapping around (4567), (6789) and (4589) directions 
of $T^6$], respectively.  Here $S_{1,\,2}^2$ are the two-cycles 
of $K3$.  
Calculations of the microscopic entropy   for such  intersecting 
$D-$brane configurations may lead to a  symmetric treatment of 
the charge degrees of freedom,  as well as to a possible 
understanding of the separate  (non-interacting) contribution 
of the left-moving and right-moving (world-volume) degrees of 
freedom.   
 
\acknowledgments

We would like to thank V. Balasubramanian, F. Larsen,  A. Sen, 
A. Tseytlin  and E. Witten for useful discussions.  The work is 
supported by the Institute for Advanced Study funds and 
J. Seward Johnson foundation, U.S. DOE Grant 
No. DOE-EY-76-02-3071, the NATO 
collaborative research grant CGR No. 940870 and the National 
Science Foundation Career Advancement Award No. PHY95-12732.

\end{document}